\documentclass[journal=jacsat,manuscript=article]{achemso}
\usepackage[version=3]{mhchem}
\usepackage{caption}
\usepackage{float}
\usepackage{setspace}
\usepackage{graphicx}
\usepackage{amsmath}
\usepackage{siunitx}

\author{Paul Weinbrenner}
\affiliation[University of Rostock]{Institute for Physics, University of Rostock, 18059 Rostock, Germany}
\alsoaffiliation{Department of Life, Light and Matter, University of Rostock, 18059 Rostock, Germany}
\author{Aina Lopez Benet}
\affiliation[University of Rostock]{Institute for Physics, University of Rostock, 18059 Rostock, Germany}
\author{İdil Gözel}
\affiliation[University of Rostock]{Institute for Physics, University of Rostock, 18059 Rostock, Germany}
\author{Friedemann Reinhard}
\email{friedemann.reinhard@uni-rostock.de}
\affiliation[University of Rostock]{Institute for Physics, University of Rostock, 18059 Rostock, Germany}
\alsoaffiliation{Department of Life, Light and Matter, University of Rostock, 18059 Rostock, Germany}
\alsoaffiliation{Munich Center for Quantum Science and Technology (MCQST), 80799 Munich, Germany}

\title[Harnessing Dielectric Interfaces]{Harnessing the Diamond-Air Interface as an Efficient Photon Antenna for Solid-State Emitters}

\abbreviations{NV, BFP, NA, ROI, SIL, QE}
\begin{document}
\begin{abstract}
Extracting photons from defect centers is challenging due to the high refractive index of typical substrates. For nitrogen-vacancy centers in diamond, reaching saturation count rates above $2.5\times10^5\,\mathrm{counts}/\mathrm{s}$ so far requires nanofabricated optics like diamond waveguides or solid immersion lenses. Here we present an experimental and theoretical study of defect center emission at unmodified planar dielectric surfaces by quantitative back focal plane imaging and analytical modeling. Our results indicate that photon count rates approaching those of nanofabricated optics can also be achieved by oil-immersion optics. This is due to a dielectric antenna effect which directs the majority of the emission into the substrate within a narrow angle window suitable for back-side collection. By quantifying the collection efficiency of back-side detection, our work also enables a novel measurement method for the quantum efficiency of shallow defect centers. Its result challenges established values. Possible reasons for the discrepancy are discussed.
\end{abstract}
\section{Introduction}
Photon extraction from defect centers is of paramount importance for quantum technologies. The sensitivity of quantum sensors~\cite{barry_sensitivity_2020} and the entanglement rate of quantum repeaters~\cite{ruf_quantum_2021} both crucially depend on the emitter's emission rate and the efficiency of photon collection. This efficiency can be increased either by non-resonant broadband collection optics~\cite{riedel_lowloss_2014, zheng_chirped_2017, kim_broadband_2025, wambold_adjointoptimized_2020, chakravarthi_inversedesigned_2020, zhu_multicone_2023} with negligible Purcell enhancement or by resonant approaches like cavities~\cite{riedel_deterministic_2017, ruf_resonant_2021, choy_enhanced_2011, hoyjensen_cavityenhanced_2020, masuda_fibertaper_2024, herrmann_coherent_2024}, which modify the density of states and achieve high Purcell factors, but are bulky and thus unsuitable for applications like sensing by near-surface defect centers.\\
For broadband photon collection from the nitrogen-vacancy (NV) center in diamond count rates in excess of $2.5\times10^5\,\mathrm{counts}/\mathrm{s}$ can only be obtained in nanofabricated optical structures like diamond solid immersion lenses (SILs)~\cite{hadden_strongly_2010, jamali_microscopic_2014, christinck_bright_2023}, diamond waveguides~\cite{momenzadeh_nanoengineered_2015} or contact-bonded high-refractive index optics~\cite{riedel_lowloss_2014}. Yet, even in these devices the photon collection efficiency is only a few percent, given that a saturated defect should emit photons at an orders of magnitude higher rate, equal to its inverse lifetime $\tau_{\text{NV}}^{-1}\approx 7\times 10^7\,\mathrm{Hz}$ ~\cite{collins_luminescence_1983, robledo_control_2010, toyli_measurement_2012, radko_determining_2016}. Why this is so and where the photons are lost is still a mystery of the field. Notably, the discrepancy cannot presently cannot presently be attributed to a poor quantum efficiency (QE) of the NV center, because it has been measured by several studies to be close to unity~\cite{gruber_scanning_1997, inam_emission_2013, mohtashami_suitability_2013, radko_determining_2016, bezard_unveiling_2024}.\\
One promising approach for non-resonant photon collection is using planar antennas made from dielectric multilayers. While these devices have not found widespread adoption for solid-state emitters, they have been studied extensively in the field of single molecules~\cite{axelrod_fluorescence_2012, chen_99_2011}, where collection efficiencies in excess of 90\,\% have been reached for molecules embedded in planar dielectric multilayer structures~\cite{leePlanarDielectricAntenna2011, chu_experimental_2014}. Similarly, the diamond surface provides a natural dielectric interface with a high refractive index contrast so that shallow NV defects natively live in a layered dielectric structure.
This motivates the key questions of this work: How exactly does the native diamond surface modify the pattern of emission? Could the diamond surface serve as a dielectric optical antenna for improved photon collection?\\
We explore this subject by two techniques: back focal plane imaging and modeling by an analytical theory of dipole emission. While both techniques have been established before and have been applied to defect centers~\cite{christinck_characterization_2020, christinck_comparison_2022}, our work moves beyond this state of the art in several important respects. First, we implement these techniques on the application-relevant case of NV centers embedded few nanometers beneath the surface of a diamond, but imaged through the diamond from its back side. In this situation, a surprisingly high photon count rate has been observed in experiments with back-side collection~\cite{weinbrenner_planar_2024}, but whether this is due to a dielectric antenna effect has remained unclear. 
Second, we perform quantitative imaging, enabling an exact quantification of the photon count rate, collection efficiency and QE which had not been possible in a previous qualitative study of the emission of NV centers close to an interface~\cite{christinck_characterization_2020}. 

\section{Results and discussion}

\subsection{Experimental setup}
\begin{figure}[ht]
\includegraphics[width=3.33in]{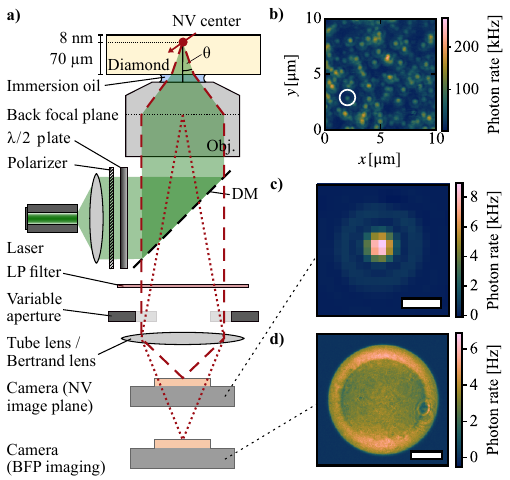}
\caption{\label{fig:setup}
Experimental setup.
\textbf{a)} Schematic of the optical setup. Obj.: Objective, DM: dichroic mirror, LP filter: longpass filter, BFP: back focal plane
\textbf{b)} Confocal photoluminescence image of the diamond sample with color-coded~\cite{crameri_misuse_2020} photon rate. The NV center used for the quantitative analysis in Fig.~\ref{fig:quantitative} is encircled in white.
\textbf{c)} Photoluminescence intensity in the image plane of a single NV center. Scale bar width corresponds to \qty{20}{\micro\metre} or $9.1$ optical units.
\textbf{d)} Image of the BFP for the emission of a single NV center. Scale bar width corresponds to \qty{400}{\micro\metre}.}
\end{figure}

The photon emission from a single NV center is studied by measuring both the emission intensity in the NV center's image plane and its angular distribution, using back focal plane (BFP) imaging/Fourier microscopy~\cite{lieb_singlemolecule_2004}. The NV center lies at a shallow depth of $\approx8\,\mathrm{nm}$ below the diamond surface (see Supplementary Information (SI) for details about sample and NV depth estimation) and the excitation and detection is performed from the bottom as shown in Fig.~\ref{fig:setup}\,a. This geometry is relevant for sensing applications~\cite{staudacher_nuclear_2013, liu_surface_2022, steinert_high_2010, pham_magnetic_2011, weinbrenner_planar_2024}, where short sensor-samples distances are crucial and optical access through the sample might be obstructed. Even through a $\approx\qty{70}{\micro\metre}$ thick diamond with high refractive index mismatch, individual NV centers remain resolvable by confocal microscopy (Fig.~\ref{fig:setup}\,b).
For excitation a $517\,\mathrm{nm}$ laser is focused with an infinity-corrected, oil-immersion objective. A polarizer and $\lambda/2$-plate control polarization, while a dichroic mirror and longpass filter separate photoluminescence from excitation.
A key requirement for the present study is seamless switching between BFP and real space imaging. The emitted light is collected by the objective, relayed by an achromatic lens, and recorded with a camera. Normal imaging is achieved when the camera is placed one focal length behind the lens, which then acts as a tube lens to project the NV center's image onto the sensor. For BFP imaging, the camera is positioned in the plane conjugate to the objective’s BFP, where the lens acts as a Bertrand lens to image the angular distribution of collected light. The effective numerical aperture (NA) of the optical setup can be modified with a variable aperture placed in front of the lens, which blocks the rays furthest away from the optical axis, corresponding to large emission angles $\theta$. To isolate the signal of a single NV center, background contributions from nearby emitters and stray light are removed by subtracting background images from the images acquired with the NV center in focus. These are acquired by shifting the excitation spot to several positions around the NV, each located at the edge of the Airy disk, i.e. first minimum of the Airy pattern of the excitation spot. For the quantitative comparison the measured count rates are divided by the camera's photon detection efficiency to obtain the actual photon rate. This value is independent of the detector's photon detection efficiency, which simplifies comparison between different detector types\\
\\

\subsection{Angular emission measurement and simulation}

\begin{figure}[ht]
\includegraphics[width=3.33in]{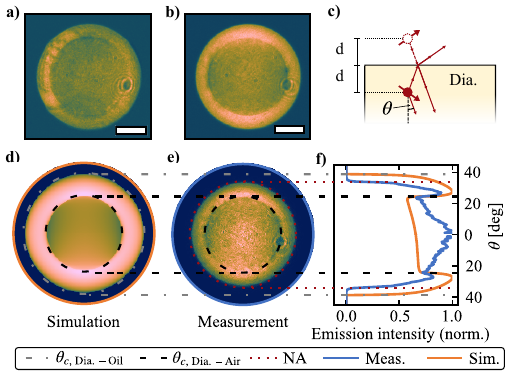}
\caption{\label{fig:bfp}
Measurement and simulation of angular emission intensity.
\textbf{a, b)} BFP images for two different NV centers. Scale bar width corresponds to \qty{400}{\micro\metre}.
\textbf{c)} Schematic of emission intensity simulation.
\textbf{d)} Simulation of the angular emission intensity. The black, dashed (gray, dash-dotted) lines indicate the angle of total internal reflection at the diamond-air (diamond-oil) interface.
\textbf{e)} BFP image converted to angular emission intensity. The red, dotted line marks the cut-off angle due to the NA of the microscope objective.
\textbf{f)} Simulated and measured angular emission intensity separately integrated over the azimuthal coordinate for the upper and lower part of the images in d, e.
}
\end{figure}

Firstly we use BFP imaging to analyze the NV center's angular emission and show that it emerges due to an interplay of three effects: the emission pattern of the NV center's electric dipoles, a dielectric antenna effect due to the proximity of the diamond-air interface, and the angular dependent attenuation from Fresnel reflection at the diamond-oil interface.\\
The BFP image (Fig.~\ref{fig:setup}\,d) displays the angular emission, where points further from the center correspond to larger emission angles $\theta$. A narrow ring with high emission intensity and sharply defined edges is apparent, forming around a central region of lower and mostly homogeneous intensity. This pattern emerges from a dielectric antenna effect of the diamond-air interface and is similar to previous results on dye molecules~\cite{lieb_singlemolecule_2004, leePlanarDielectricAntenna2011, checcucciBeamingLightQuantum2017}, quantum dots~\cite{christinck_comparison_2022} and emitters in nanodiamonds~\cite{christinck_characterization_2020} imaged through transparent substrates. 
Along the vertical axis through the image center, the intensity is enhanced compared to the lateral directions. The pattern is asymmetric, because the inner edge of the ring has a higher contrast on the top side than on the bottom. Comparing the BFP images of different NV centers (Fig.~\ref{fig:setup}\,d and Fig.~\ref{fig:bfp}\,a, b) reveals the same features, only rotated in steps of $90^\circ$. \\
We explain these features by the crystallographic orientation of the NV center and the resulting orientation of its optical dipoles, which lie in the plane perpendicular to its symmetry axis~\cite{epstein_anisotropic_2005} and define the angular emission pattern~\cite{lieb_singlemolecule_2004, riedel_lowloss_2014, christinck_characterization_2020}. 
Simulations suggest that the vertical or horizontal emission peaks correspond to a pair of NV orientations oriented horizontally or vertically on our $\left(100\right)$-polished diamond surface and that the two orientations within one of these pairs differ by the slight asymmetry within one family of BFP patterns. This interpretation is confirmed by polarization dependent photoluminescence measurements (see SI). As a result, the orientation of a NV center can be identified from a single BFP image.
\\
Remarkably, the essential features of the BFP image can be captured by a simple analytical theory model~\cite{lukosz_light_1981, lethiec_measurement_2014, riedel_lowloss_2014}, illustrated in Fig.~\ref{fig:bfp}\,c, which superimposes the direct downward emission of the NV center's electric dipoles with the part of the upward emission which is reflected at the diamond-air interface. Effectively, the total field is obtained by superimposing the dipole’s electric field with that of a virtual source above the diamond surface, weighted by the Fresnel reflection coefficients. Transmission through the diamond–oil interface is then included via the corresponding Fresnel coefficients.\\
The dielectric antenna effect becomes evident when plotting the angular emission in polar coordinates, where the radius represents the emission angle $\theta$ inside the diamond. The sharp intensity increase coincides with the critical angle for total internal reflection at the diamond–air interface $\theta_{c,\mathrm{Dia.-Air}}=24.5^\circ$. This additional light corresponds to photons emitted into the upper half-space that undergo total internal reflection at the diamond surface. Beyond this intensity peak the emission decreases due to partial reflection at the diamond-oil interface until it vanishes at its critical angle $\theta_{c,\mathrm{Dia.-Oil}}=38.8^\circ$.\\
A comparison with the measured angular emission in Fig.~\ref{fig:bfp}\,e shows that the simulation reproduces the main features, i.e. the dipole emission pattern and high intensity ring. The measured intensity is cut off before $\theta$ reaches the critical angle $\theta_{c,\mathrm{Dia.-Oil}}$ because of the finite NA of the microscope objective. Despite the overall good agreement, some discrepancies between simulation and measurement are apparent, especially in the azimuthally averaged emission intensity in Fig.~\ref{fig:bfp}\,f. The measured angular emission has a higher intensity at the center and decreases faster for large $\theta$ than predicted by the simulation. Reasons for this might be vignetting in our optical setup and the microscope objective's deviation from the ideal apodization curve~\cite{kurvits_comparative_2015}. Additionally, the simulation does not reproduce the bright, central spot in the BFP which we see on few, specific NV centers (Fig.~\ref{fig:quantitative}\,h), which we presently cannot explain and merely note as an observation.
\subsection{Image plane analysis and spherical aberrations}
\begin{figure}[ht]
\includegraphics[width=3.33in]{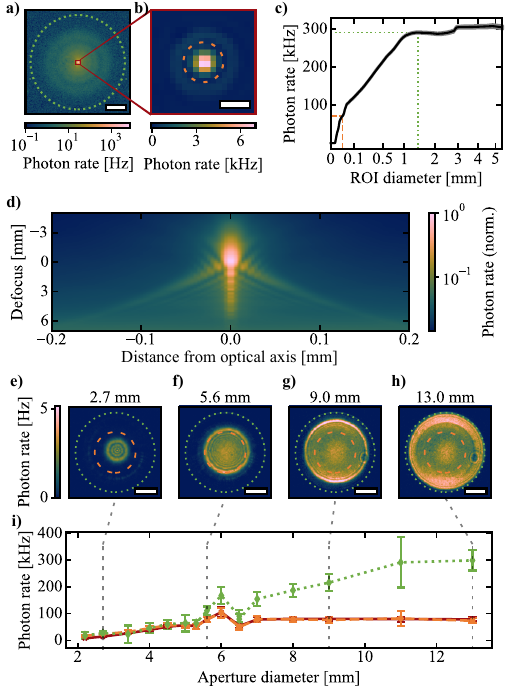}
\caption{\label{fig:quantitative}
Quantitative photon rate analysis in image plane and BFP.
\textbf{a)} Emission intensity in NV center image plane with logarithmic color scale. The green, dotted outline indicates region of interest (ROI) with $1.4\,\mathrm{mm}$ diameter. Scale bar width corresponds to \qty{300}{\micro\metre}.
\textbf{b)} Zoom into the image in a), but with linear color scale. The circular ROI with yellow, dashed outline shows the edge of the Airy disk with \qty{27}{\micro\metre} diameter. Scale bar width corresponds to \qty{20}{\micro\metre}.
\textbf{c)} Unsaturated photon rate within circular ROIs with different diameter in the NV center image plane with the uncertainty illustrated by the shaded region. The yellow, dashed (green, dotted) line marks the photon rate for a \qty{27}{\micro\metre} ($1.4\,\mathrm{mm}$) ROI (cf. a and b). The x axis is scaled quadratically.
\textbf{d)} Azimuthal average of the photon rate when defocusing the detection plane from the pinhole plane.
\textbf{e)-h)} Photon rate in the BFP for different diameters of the variable aperture. Scale bar width corresponds to \qty{400}{\micro\metre}.
\textbf{i)} Photon rate as a function of the diameter of the variable aperture. Yellow, dashed line: photon rate for the inner ROI in the BFP, as indicated in e)-h). Green, dotted line: total photon rate in the BFP, as indicated in e)-h). Red, solid line: total photon rate within the Airy disk, measured with confocal setup (see SI), value corresponds to intensity encircled in the ROI in b).
Note: The stated photon rates are for an unsaturated NV center and correspond to count rates for a detector with unity photon detection efficiency. See the main text for the saturation count rate achievable with typical single photon detectors.
}
\end{figure}

We now turn to the central question of this work: given that a strong dielectric antenna effect appears at the diamond-air interface, what implications does this have for the count rate obtained from a single NV center under back-side collection in a standard oil-immersion confocal microscope? We address this question by quantitatively analyzing the emission of a single NV center in the image plane and the BFP (Fig.~\ref{fig:quantitative}).\\
The NV center's image is an Airy pattern with an unusually bright and broad second maximum surrounding the central peak (Fig.~\ref{fig:quantitative}\,b). When zooming out of the central focal spot (Fig.~\ref{fig:quantitative}\,a) a broad intensity distribution around it can be observed, indicating that a substantial fraction of photons are not focused into the diffraction-limited spot. This results from strong spherical aberrations induced by the large refractive index mismatch of the \qty{70}{\micro\metre} thick diamond that the NV center is imaged through~\cite{booth_aberration_1998}. In the defocused images of the NV center emission in Fig.~\ref{fig:quantitative}\,d, the asymmetry around the focal plane (Defocus$\,=0\,\mathrm{mm}$) and strong intensity at the edges of the Airy pattern are also telltale signs of spherical aberrations.
Diamond-adapted optics could mitigate these aberrations, but are not commercially available today.\\ 
\\
Crucially, we can assess what photon rate could be obtained by such optics, since our imaging is quantitative. For an exact quantification, we integrate the detected photon rate within circular regions of interest (ROIs) of increasing diameter (Fig.~\ref{fig:quantitative}\,c).
The encircled photon rate increases steeply for small ROI diameters less than \qty{50}{\micro\metre}, because of the high intensity of the Airy disk and the surrounding halo. However, photon collection continues to rise up to a ROI diameter of $1.4\,\mathrm{mm}$, far exceeding the detection area typically used in microscopy. In a conventional confocal microscope the pinhole size is chosen so that only the Airy disk is detected, corresponding to a diameter of \qty{27}{\micro\metre} in this work (yellow, dashed line in Fig.~\ref{fig:quantitative}\,b).
Restricting the detection to this region yields a photon rate of $71.0\pm1.0\,\mathrm{kHz}$. Opening the aperture to the full $1.4\,\mathrm{mm}$ increases the rate more than fourfold to $291\pm3\,\mathrm{kHz}$. Correcting for the fact that this measurement has been performed at an excitation power far from saturation (see SI), this corresponds to a saturation photon rate of $511\pm30\,\mathrm{kHz}$. A typical avalanche photo diode with a photon detection efficiency of $68\%$ would thus detect a saturated count rate of $346\pm20\,\mathrm{kcounts/s}$, if the light could be focused onto the active detection area. This significantly surpasses detection from the top-side and approaches SIL-~\cite{hadden_strongly_2010, jamali_microscopic_2014} and waveguide-level~\cite{momenzadeh_nanoengineered_2015} count rates. This photon rate is consistent with a similar estimate from the BFP image in Fig.~\ref{fig:quantitative}\,h. Integrating the BFP intensity also predicts an unsaturated photon rate of $299\pm38\,\mathrm{kHz}$.
The enhancement of photon collection is further validated by a measurement of the photon emission on the same NV center with oil-immersion detection from the top-side (see SI), where we find a saturation photon rate of only $275\pm13\,\mathrm{kHz}$ inside the $1.4\,\mathrm{mm}$ ROI and $192\pm10\,\mathrm{kHz}$ within the \qty{27}{\micro\metre} diameter Airy disk.
\\
Analyzing the BFP emission furthermore allows a study of how exactly photons are lost to spherical aberrations, which indicates that only light emitted under small angles is focused onto a pinhole-sized area for easy detection, which is presented in detail in Fig.~\ref{fig:quantitative}\,e-i. We gradually open an aperture in front of the Bertrand/tube lens, so that high-angle rays are progressively included in the detection path. At each aperture setting, we record both the total photon rate in the BFP and the photon rate within the Airy disk of the image plane (Fig.~\ref{fig:quantitative}\,i). The latter is measured with a confocal detection path (see SI) that is equivalent to integrating the photon rate within a ROI of \qty{27}{\micro\metre} diameter on the camera as shown in Fig.~\ref{fig:quantitative}\,b. Both signals increase equally as the aperture opens from the fully closed position. However, once the diameter exceeds $6\,\mathrm{mm}$ (dashed ring in Figs.~\ref{fig:quantitative}\,e-h), corresponding to an emission angle of $28^\circ$ or NA of $0.72$, the Airy disk photon rate saturates, while the BFP intensity continues to rise. This indicates that beyond this aperture diameter only light emitted under larger angles is added to the detection path, which is not focused inside the Airy disk, because of spherical aberrations, and thus would be rejected by a pinhole in a confocal detection. In summary, spherical aberrations prevent full exploitation of the dielectric antenna effect and reduce the effectively used NA of oil-immersion objectives to $0.72$.
\subsection{Collection efficiency}
\begin{figure}[ht]
\includegraphics[width=3.33in]{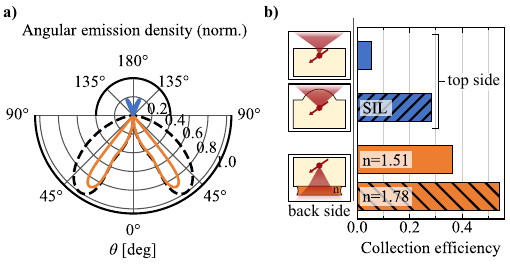}
\caption{\label{fig:colleff}
Simulation of the photon collection efficiency.
\textbf{a)} Angular emission density. The blue line on the upper part shows the emission intensity towards the top, through the diamond-air interface. The yellow (black, dashed) line show the emission towards the back side including (excluding) the Fresnel transmission coefficients at the diamond-oil interface.
\textbf{b)} Collection efficiency for different detection geometries. For oil immersion with $n=1.51$ ($n=1.78$) a numerical aperture of $1.45$ ($1.7$) was used.
}
\end{figure}

%The surprising conclusion that solid-immersion-level count rates could be obtained by back-side immersion optics is supported by a computation of the collection efficiency for different detection geometries presented in Fig.~\ref{fig:colleff}.
The significant increase in photon rate by using back-side detection is supported by a computation of the collection efficiency for different detection geometries presented in Fig.~\ref{fig:colleff}.
In addition to the previous emission pattern simulation for collection from the back side of the diamond, we calculate the emitter's angular emission into the upper half-space (air), where no dielectric antenna effect occurs.
The total intensity emitted upwards is thus given by the electric dipoles' emission pattern weighted by the Fresnel transmission coefficients at the diamond-air interface. Fig.~\ref{fig:colleff}\,a shows the resulting angular emission density, integrated over the azimuthal coordinate, together with the corresponding emission towards the substrate side, both with and without Fresnel transmission at the diamond–oil interface.
\\
Emission into air is strongly suppressed due to the large refractive index mismatch. The low critical angle at the diamond–air interface restricts transmission to a narrow angular range and even within this angle the Fresnel reflection coefficient remains high. Thus, much of the emission is reflected back into the diamond and redirected towards the substrate by the dielectric antenna effect. While the emission into the bottom half-space is much stronger, it is itself limited by total internal reflection at the diamond–oil interface, which allows only $\approx40\%$ of the downward emission to exit the diamond.\\
To assess realistic collection efficiencies, we calculate the detectable light intensity and divide it by the total emission for four geometries: (i) top-side collection through air, (ii) top-side collection with a SIL, (iii) back-side collection with standard oil immersion ($n=1.51$), and (iv) back-side collection with high-index oil ($n=1.78$). Notably, the total emission intensity is different for geometry (ii), because of the different dielectric environment of the emitter (see SI).
In all cases the finite acceptance of the objective is taken into account by truncating the integration at the corresponding NA-angle $\theta_\mathrm{max}=\arcsin\left(\mathit{NA}/n\right)$ with $\mathit{NA}=0.95$ for detection from the top side through air, $\mathit{NA}=1.45$ and $\mathit{NA}=1.7$ for the collection from the diamond's back side with normal and high-$n$ immersion oil respectively.
The collection efficiencies are summarized in Fig.~\ref{fig:colleff}\,b. For top-side emission the high refractive index mismatch mentioned above results in a collection efficiency of only $5.4\%$.
Using a diamond SIL circumvents total internal reflection, raising the collection efficiency to $28\%$, consistent with prior reports~\cite{hadden_strongly_2010, jamali_microscopic_2014, christinck_bright_2023}.
Surprisingly, similar and even higher efficiencies are achieved without nanofabrication: back-side collection with standard immersion oil reaches $36\%$, and with high refractive index oil increases further to $54\%$, which is a $93\%$ improvement over diamond SILs.\\
Intriguingly, the ability to exactly quantify the collection efficiency and photon rate also allows for an independent measurement of the NV center's QE $\eta$
\begin{equation}
\eta=\frac{I_{\mathrm{det}}}{\tau_{\mathrm{NV}}^{-1}\,\mathit{CE}\,T_{\mathrm{setup}}\,p_{\mathrm{det}}}~,
\end{equation}
where $I_{\mathrm{det}}$ is the detector count rate in saturation, $\tau_{\mathrm{NV}}=11.6\,\mathrm{ns}$ is the excited state lifetime~\cite{collins_luminescence_1983}, $\mathit{CE}$ is the collection efficiency, $T_{\mathrm{setup}}=74\%$ is the transmission coefficient of the optical setup (see SI) and $p_{\mathrm{det}}$ is the detector's photon detection efficiency~\cite{lakowicz_introduction_2006}.
The resulting value $\eta=2\%$ is much lower than previously reported quantum efficiencies~\cite{gruber_scanning_1997, inam_emission_2013, mohtashami_suitability_2013, radko_determining_2016, bezard_unveiling_2024} and more in line with related estimates of the silicon-vacancy center's quantum efficiency~\cite{neu_photophysics_2012}.
One possible reason for this discrepancy could be vignetting in the optical setup, since especially high-$\theta$ rays undergo strong aberrations and might be clipped even within the microscope objective. 
Furthermore, our study operates on $\approx8\,\mathrm{nm}$ shallow NV centers, which are known for center-to-center and diamond-to-diamond variability in their physical properties. We might have picked a diamond or NV center with a low QE. Also, our study does not correct for NV$^-$/NV$^0$ charge state switching and might thus underestimate the NV$^-$ count rate.
Finally, our theory model neglects multi-path reflections at the diamond's bottom and top surfaces. These could develop into lateral waveguide modes, that elude collection by the objective and hence provide an additional decay channel, invisible to the methods of our study.
However, we believe that, with the exception of waveguide modes, these explanations are unlikely to account for a discrepancy of this scale and that several other facts call for a further study and possibly reconsideration of the NV center's QE.
First, recent measurements that all found QEs larger than $60\%$ hinge on specific assumptions about the ionization rate from NV$^-$ to the neutral charge state~\cite{radko_determining_2016}, the modeling of the local density of states for nanodiamonds~\cite{bezard_unveiling_2024} or the QE of deep NV centers~\cite{inam_emission_2013}. A widely cited reference stating unity QE assumes the singlet state as the only possible non-radiative relaxation pathway~\cite{rand_diamond_1994}.
Second, photon rates on the order of the inverse lifetime $\tau^{-1}\approx 5\times10^7\,\mathrm{Hz}$ have indeed been harvested from single terrylene molecules in a geometry that is closely related to our study~\cite{leePlanarDielectricAntenna2011}.
However, assumptions of unity quantum efficiency have also shown good agreement with measurements of the NV center emission inside a resonant microcavity~\cite{riedel_deterministic_2017}.
Intriguingly, our study paves the way to resolve this disagreement. Modifying the local density of states by exchanging the medium on the diamond surface could provide an independent measurement of the NV center's QE in this work. Placing a high-QE emitter like terrylene in the same geometry and setup next to a NV center on the diamond surface could provide a quantitative one-to-one comparison.
\subsection{Conclusion and outlook}
This work quantitatively studies the emission of emitters few nanometers beneath an unmodified diamond-air interface, and the strong dielectric antenna effect that occurs in this setting. The results open several technological and scientific perspectives.\\
First, back-side collection with oil-immersion (or glass-based solid-immersion optics) can achieve saturation count rates significantly higher than standard top-side collection , i.e. $346\pm20\,\mathrm{kcounts/s}$ for a single NV center in this study. 
The collection efficiency of back-side collection can even exceed nanofabricated diamond optics ($54\%$ with high refractive index oil compared to $28\%$ for an all-diamond SIL).
This insight provides a strong motivation to design correction optics which mitigate diamond-induced spherical aberrations and actually allow harvesting these photon count rates in a confocal setup.\\
Second, count rates comparable to the state of the art in through-diamond, back-side photon collection can likely be obtained by much simpler objective lenses, e.g. oil immersion objectives of $\mathit{NA}=0.72$, since the high-NA lenses currently employed only make use of a small fraction of their nominal NA due to spherical aberrations.\\
Third, combining our analytical estimate of the collection efficiency with a quantitative analysis of the measured photon rate provides an independent route to measuring the QE, which suggests a much lower QE for NV centers than previously reported. It raises questions about the intrinsic emission dynamics of near-surface NV centers, but also paves a way towards finally resolving them.\\
Taken together, our results identify the dielectric antenna effect as a powerful method for efficient photon collection not only for NV centers in diamond, but also other solid state emitters.
\subsection{Author contributions}
P.W. and F.R. conceived the project. A.L.B., İ.G. and P.W. built the experimental setup. P.W. performed the experiment under supervision from F.R.. P.W. and F.R. created the simulation, analyzed the data and wrote the paper. All authors read and commented on the final manuscript.

\begin{acknowledgement}
This work was supported by the SFB 1477 “Light-Matter Interactions at Interfaces” (Project No. 441234705). A.L.B. acknowledges support from DAAD Rise Germany and İ.G. acknowledges support from Erasmus+. The authors thank Thomas Fennel for helpful discussions.
\end{acknowledgement}

\begin{suppinfo}
Source code for emission simulation; Supplementary Information on: Diamond substrate and NV center creation, Optical setup, Quantitative comparison of camera and confocal detection, Photon detection efficiency and transmission of optical setup, Polarization dependent photoluminescence, Transformation from BFP image to angular emission intensity, Collection efficiency calculation, Depth dependence of collection efficiency, Saturation photon rate measurement, Comparison of back-side and top-side detection
\end{suppinfo}

\subsection{Competing interests}
The authors declare no competing interests.

\bibliography{NV_photon_emission.bib}

\end{document}